\documentclass[12pt,preprint]{aastex}
\usepackage{graphicx}
\newcommand{\Msun}{\rm{M}_\odot}
\title{Possible Molecular Spiral Arms in the Protoplanetary Disk of AB Aur}
\author{
Shin-Yi Lin\altaffilmark{1, 2}, Nagayoshi Ohashi\altaffilmark{2},
Jeremy Lim\altaffilmark{2}, Paul T. P. Ho\altaffilmark{2,3}, Misato
Fukagawa\altaffilmark{4}, and Motohide Tamura\altaffilmark{4}}
\altaffiltext{1}{Institute of Astronomy, National Tsing Hua
University, 101, Section 2 Kuang Fu Road, Hsinchu 30013, Taiwan}
\altaffiltext{2}{Academia Sinica Institute of Astronomy and
Astrophysics, P.O. Box 23-141, Taipei 106, Taiwan}
\altaffiltext{3}{Harvard-Smithsonian Center for Astrophysics, 60
Garden Street, Cambridge, Massachusetts 02138, USA}
\altaffiltext{4}{National Astronomical Observatories of Japan, Osawa
2-21-1, Mitaka, Tokyo 181-8588, Japan}
\email{sylin@asiaa.sinica.edu.tw}
\begin{abstract}
The circumstellar dust disk of the Herbig Ae star AB Aur has been found to exhibit complex spiral-like
structures in the near-IR image obtained with the Subaru Telescope. We present maps of the disk in both
$^{12}$CO (3-2) and dust continuum at 345~GHz with the SMA at an angular resolution of $1''\!\!.0$ $\times$
$0.7''\!\!$~(144AU $\times$ 100AU). The continuum emission traces a dust disk with a central depression and a
maximum overall dimension of 450AU (FWHM). This dust disk exhibits several distinct peaks that appear to
coincide with bright features in the near-IR image, in particular the brightest inner spiral arm. The CO
emission traces a rotating gas disk of size 530AU $\times$ 330AU with a deprojected maximum velocity of
2.8~km~s$^{-1}$ at 450~AU. In contrast to the dust disk, the gas disk exhibits an intensity peak at the stellar
position. Furthermore, the CO emission in several velocity channels traces the innermost spiral arm seen in the
near-IR. We compare the observed spatial-kinematic structure of the CO emission to a simple model of a disk in
Keplerian rotation, and find that only the emission tracing the main spiral arm clearly lies outside the
confines of our model. This emission has a net outward radial motion compared with the radial velocity predicted
by the model at the location of the main spiral arms. The disk of AB Aur is therefore quite different from the
Keplerian disks seen around many Herbig Ae stars. The spiral-like structures of the disk with non-Keplerian
motions we revealed in $^{12}$CO(3-2), together with the central depression of the dust disk, may be explained
to be driven by the possible existence of a giant planet forming in the disk.

\end{abstract}
\begin{document}
\maketitle
\section{Introduction}
Circumstellar disks of low and intermediate mass pre--main-sequence stars, namely T Tauri stars and Herbig Ae/Be
stars, respectively, are generally considered to be the birth places of planetary systems
\citep[cf.][]{Zuckerman:2001}. Knowledge of the disk structure and kinematics is vital for understanding the
physical environment of planet formation. Herbig Ae stars, in particular, are considered to be the progenitors
of prototypical Vega-like stars, which are surrounded by debris-disks. These disks are believed to be produced
by colliding planetesimals as well as perhaps planets.  How does a circumstellar disk containing gas and dust
evolve into a disk containing primarily dust? By observing the evolving circumstellar disks of Herbig Ae stars
and looking for evidence for actual formation of (proto)planets, we hope to provide instructive clues to the
process of planet formation. This work may provide a snapshot at or close to the starting point.

\begin{figure}
\begin{center}
\includegraphics[width = .45\textwidth]{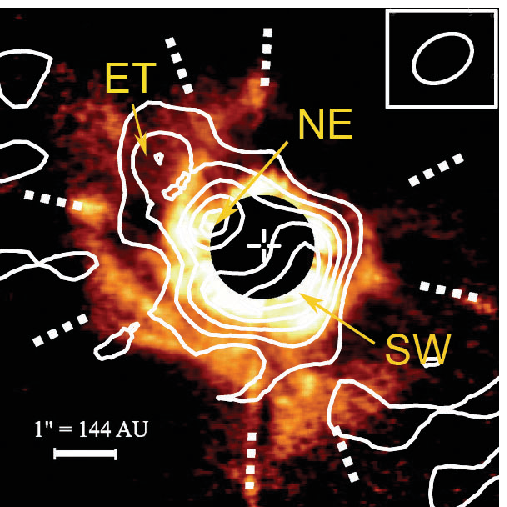}
\includegraphics[width = .45\textwidth]{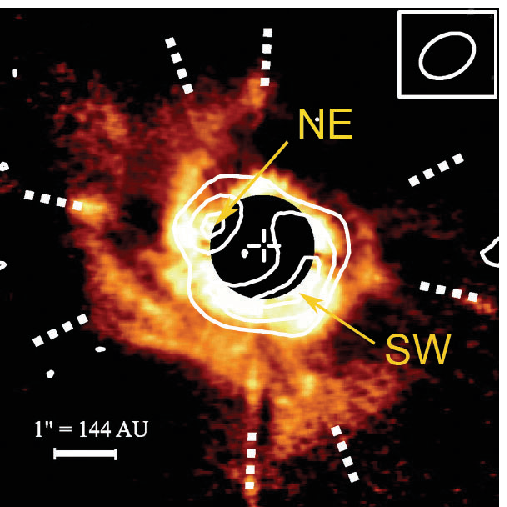}
\caption[Dust continuum maps]{Left: Dust continuum with natural
weighting superposed on the Subaru near-IR image. The angular
resolution is $1''\!\!.04\times~0''\!\!.72$. Contours start from
2$\sigma$ (or 5.5mJy beam $^{-1}$) with a spacing of 2$\sigma$.
Right: The dust continuum without inner 30k$\lambda$ data points on
the UV plane superposed on the Subaru image. The angular resolution
is $0''\!\!.95\times~0''\!\!.66$. Contour spacing is 2$\sigma$ (or
5.6mJy beam$^{-1}$). Note that the central part of the near-IR image
has been suppressed with an occulting mask. The SMA dust image shows
a definite depression in the center. } \label{Fig 1}
\end{center}
\end{figure}

AB Aur \citep[V = 7.06 $\pm$ 0.06, spectral type = A0~Ve + sh, d=144pc,][]{vandenAncker:1997} is one of the
nearest and best studied Herbig Ae stars. With an age of 2 - 5 Myr \citep{DeWarf:2003}, it not only has a bright
dust disk, but is also rich in gas. It is thus an ideal place to study the initial condition of planet
formation. The moderate inclination angle also makes it a desirable source for imaging. Previous millimeter
wavelength observations with an angular resolution of 5$''$ made by \cite{Mannings:1997} with the OVRO
interferometer in CO~(1-0) revealed a velocity gradient along the major axis attributed to a disk in Keplerian
motion. Optical imaging with the HST \citep{Grady:1999} showed a north-south asymmetry in the dust disk. Near-IR
imaging with the Coronographic Imager with Adaptive Optics on the Subaru Telescope shows an even more
complicated structure \citep[hereafter F04, see Figure \ref{Fig 1}]{Fukagawa:2004}. F04 identified at least two
apparent spiral-like structures in the dust disk: a prominent inner arm with a radius of 230 AU from the east to
northeast, and an outer arm with a radius of 330 AU from the south to northeast. We will use ``the inner arm''
and ``the outer arm'' to refer to these two arms respectively hereafter. AB Aur is one of the few stars around
which prominent spiral structures have been discovered in the circumstellar dust \citep[c.f.~][]{Clampin:2003}.
However, the near-IR image traces scattered light from the dust disk, and provides no information beyond the
surface of the optically thick dust disk. The apparent spiral structures may just be surface features, and the
existence of density spiral arms need to be verified. It would also be intriguing to measure the kinematics of
the arms and to look for streaming motion, if any, along or across the spirals to examine the possibility of
dynamical perturbations from underlying planetary bodies.

Imaging at millimeter or sub-millimeter wavelengths with higher angular resolution is able to trace the dust and
molecular gas emission directly from the disk, allowing us to study both its structure and kinematics in greater
detail. Since the best millimeter images available at the time we started our project was the 5$''$ OVRO
observation \citep{Mannings:1997}, we surmised that the complex disk morphology expected from the Subaru image
were smoothed by the large synthesized beam. Here, we present submillimeter imagings of the dust and gas disk of
AB Aur at a wavelength of 850 $\micron$ with the Submillimeter Array \citep[SMA\footnote{The Submillimeter Array
(SMA) is a joint project between the Smithsonian Astrophysical Observatory and the Academia Sinica Institute of
Astronomy and Astrophysics, and is funded by the Smithsonian Institution and the Academia Sinica.},][]{Ho:2004},
with an angular resolution of 1$''$ to unveil the hidden information.

Millimeter observations conducted at the same time as this work, by \cite{Corder:2005} and Pi\'etu et al.~(2005,
hereafter P05) with comparable angular resolutions, revealed interesting results. \cite{Corder:2005}\ obtained
images of the $^{13}$CO(1-0) line and 2.7 millimeter continuum emission at an angular resolution of $\sim$2$''$,
and images of $^{12}$CO(1-0) and C$^{18}$O(1-0) line emission at lower angular resolutions with the OVRO. After
subtracting a best-fit Gaussian source from their 2.7 mm dust continuum map, Corder et al.~see weak residuals
that appear to roughly coincide with the spiral arms seen in the near-IR. With 2$''$ angular resolution, the
$^{13}$CO(1-0) emission is satisfactorily fit by a disk with radius of $\sim615$ AU in Keplerian rotation, as
the early OVRO observation had claimed \citep{Mannings:1997}. Their $^{12}$CO(2-1) map observed at $\sim2''$,
however, can not be fit by a pure-disk model. Corder et al.~attributed the discrepancy to the contamination of a
large scale envelope. \cite{Pietu:2005}, observed the circumstellar disk of AB Aur with the PdBI in
$^{13}$CO(2-1) and in continuum at 1.4 mm with the highest angular resolution currently available
($\sim0''\!\!.59$), as well as other CO isotopes at lower resolution. They discovered asymmetric distribution of
dust continuum emission, which follows inwards along the spiral-like features. However, their results of the
$\chi^2$-fittings of the $^{13}$CO gas kinematics showed a velocity profile shallower than Keplerian. On the
other hand, the $^{12}$CO velocity law is much steeper than Keplerian. Pi\'etu et al.~also interpreted the
non-Keplerian kinematics as the results of contamination from the extended envelope but not from the disk
itself.

In contrast with the contemporary OVRO and PdBI observations, our work presents the highest resolution images of
a higher transitional line ($J=3-2$) of $^{12}$CO. Because AB Aur is still embedded in a relatively extended
envelope \citep[e.g.~][]{Grady:1999, Semenov:2005}, observations of molecular gas at its higher transitions
(which, for the same molecule, traces denser and/or warmer gas), should suffer less contamination from the
surrounding envelope and therefore better trace the disk. Moreover, flux density measurement at high frequency
extends the calculation of the spectral index to the sub-mm regime. With the capability of the SMA to observe at
higher frequency and high angular resolution, our goal is to unveil the hidden structure and kinematics of the
possible spiral arms.

\section{Observations}
We observed AB Aur over three different tracks in the dust continuum
at 345 GHz and the $^{12}$CO (3-2) rotational transition
simultaneously with the SMA. The observations were made over the
period from Dec 2003 to Sep 2004, with one compact and two extended
array configurations. The projected shortest and longest baselines
are $\sim$14 m and $\sim$220 m, respectively. Data obtained from
seven of the eight antennae were usable. The resultant size of the
synthesized beam was $1''\!\!.0\times~0''\!\!.72$ (144AU $\times$
115AU) with natural weighting (P.A.$\sim$ -58$^{\circ}$). The
correlater was configured to provide a narrow band of 512 channels
over 104 MHz and hence a velocity resolution of 0.17 km s$^{-1}$ for
the CO(3-2) line, and an overall bandwidth of 2 GHz in each of the
lower and upper sidebands for measuring the dust continuum from the
line-free channels.

Bandpass calibration that calibrates the instrumental response of
different velocity channels was done with Jupiter and quasar
2232+117 at different observing epochs. Flux calibration was made
with Uranus and the moon of Jupiter, Callisto. Amplitude and phase
calibration was done with either the strong quasar 0359+509 or 3C84
on different observation epochs, which have an angular separation of
21.7$^{\circ}$ and 26.4$^{\circ}$ from AB Aur, respectively. We used
either the weaker quasar 3C 111 or 3C 120, with angular separation
of 10.7$^{\circ}$ and 25.6$^{\circ}$, respectively, from AB Aur, to
verify the quality of the phase referencing from both 0359+509 and
3C 84. The maps of both 3C111 and 3C120 look like point sources at
phase center, as would be expected for proper phase referencing,
verifying that our positions are accurate to 0$''\!\!.1$. The same
calibration scheme was applied to AB Aur. We used the software MIR,
adapted for the SMA from an IDL-based package originally developed
for the OVRO array \citep{Scoville:1993}, for calibrating the data.
We then used the NRAO AIPS package to produce images of both the
continuum and molecular line.

\section{Results}
\subsection{Dust Continuum Emission}
We measured a total flux density of 235 $\pm$ 42mJy for the dust
continuum emission within a radius of $\sim$350AU from the central
star. This is about 65$\%$ of the flux density observed by the JCMT
\citep{Mannings:1994}. It is not surprising since the JCMT is
sensitive to the extended envelope as well as the circumstellar disk
while the extended envelope was resolved out by the SMA. Note that
the JCMT primary beam is even smaller than the SMA primary beam,
which implies that the missing flux may be a lower limit.

The dust continuum map made with natural weighting is superposed on the Subaru near-IR image in the left panel
of Figure 1. The continuum emission has an overall size of 450AU $\times$ 270AU (P.A.~66$^{\circ}$), derived by
a Gaussian fit to the data of the most compact array configuration, which only marginally resolved the disk. The
emission does not peak at the stellar optical position, nor does it exhibit an intensity distribution decreasing
monotonically in the radial direction, as is generally found for circumstellar disks associated with
pre--main-sequence stars. Instead, the disk shows two distinct peaks - one the northeast direction (NE) and the
other southwest (SW) - and an extension towards the northeast (labeled ET in Figure 1). The two peaks form a
ring-like structure with a radius of 150 AU (measured at the maximum of the two peaks) centered at the stellar
position. They also coincide with the inner region of the prominent inner arm inferred by F04 from the Subaru
image, as suggested by P05. Compared with the 1.4 mm continuum image of the dust disk made by P05, we find that
the SW peak is in agreement with the local column density enhancement $\sim$1$''$ west from the center in their
1.4 mm map, while there is no apparent peak (no local maxima) in that map at the NE position.

The northeast extension ET, which does not have a counterpart in the
PdBI map, appears to be closely aligned with the outer region of the
same spiral arm traced by the NE and SE peaks closer to the star.
Because of missing short spacings with the SMA, large scale
structures are not well sampled, which produces large-scale
corrugated features at low levels in our map. We suppressed this
effect by removing data points with uv distances smaller than
30k$\lambda$. The right panel of Figure \ref{Fig 1} presents the
resultant map superposed on the Subaru image. The northeast
extension, which was a 5$\sigma$ detection, disappears in this
image, suggesting that the ET feature is possibly part of a large
scale disk or envelope. On the other hand, this extension is close
to our CO spiral arms (see Section 3.2), suggesting that it may be
an enhancement in a larger-scale disk, which is difficult to image
against possible enhancements in a large-scale envelope.

\subsection{Molecular Gas}
$^{12}$CO (3-2) traces warmer and denser regions of the molecular gas disk as compared to the lower CO
transitions. The JCMT observations of $^{12}$CO(3-2)/$^{13}$CO(3-2) line ratio by \cite{Thi:2001} show that the
optical depth of $^{13}$CO is 0.21. Assuming a $^{12}$C/$^{13}$C isotopic ratio of 60, we find that the
$^{12}$CO emission is optically thick, and will likely trace the gas temperature distribution close to the disk
surface (c.f.~P05). In contrast, the dust emission at 850 $\micron$ is likely to be optically thin and hence a
good tracer of column density. The measurements of dust and CO emission are complementary, and together they
provide a more complete picture of the properties of the disk. Note that the optical depth computed here with
the observation of Thi et al.~is an average of the disk and any envelope component, since the observation was
done by single dish telescope with a resolution of $\sim$ 14''. The computed high optical depth does not imply
that the disk is optically thick everywhere.  At some locations, either or both the column density and
excitation temperature may not be high enough to make $^{12}$CO (3-2) optically thick.

\begin{figure}
\begin{center}
\includegraphics[width=\textwidth]{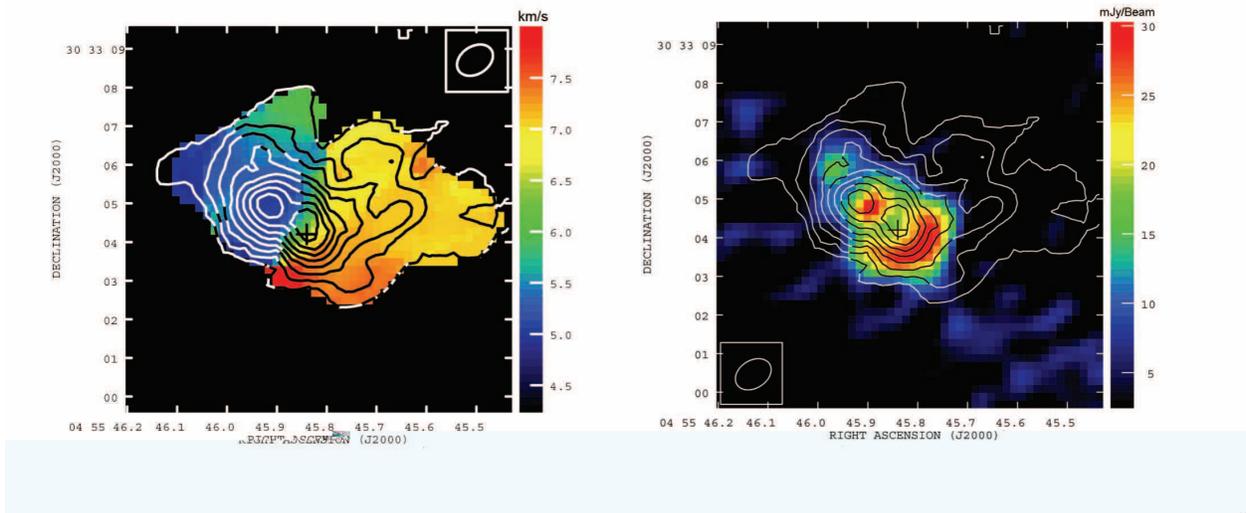}
\caption[CO moment maps]{Left: CO moment maps. The contours
represent the CO integrated intensity (zeroth moment), with 0.2Jy
Beam$^{-1}$ $\times$ km s$^{-1}$ (1$\sigma$) spacing. Here the
``noise'' of the final integrated map was estimated from integrating
the r.m.s.~noise of 6 channels with a bandwidth of $\sim$1km
s$^{-1}$, because at any position the line width is smaller than 1km
s$^{-1}$. The color scale map is the mean velocity (first moment)
map from 4.2km s$^{-1}$ to 8.0km s$^{-1}$. Right: CO MOM0 map
superposed on dust continuum map. The off-center peak of the CO is
very close to the NE peak of the dust continuum emission. Crosses in
both panels represent the stellar position. \label{Fig 3} }
\end{center}
\end{figure}

We detected significant emission ($>$3$\sigma$) in velocity channels between 4.1 km s$^{-1}$ to 8.0 km s$^{-1}$
with a peak brightness temperature of 34.3 $\pm$ 4.0K. We only detect 20\% of the CO(3-2) flux measured by the
JCMT \citep{Thi:2001}, suggesting that the single-dish observation contains considerable emission contributed
from the large scale envelope. The CO line emission did not suffer from the large-scale ripple effect as the
dust emission, perhaps because in a given velocity channel the CO emission is not as extended as the dust.
Therefore the inner 30k$\lambda$ data were retained for maximum sensitivity. The left panel of Figure \ref{Fig
3} shows the integrated intensity map of CO over the velocity channels from 4.1 km s$^{-1}$ to 8.0 km s$^{-1}$,
superposed on the mean velocity map in false color. These maps were made using "MOMNT" in AIPS to bring out
non-random signals and suppress random noise fluctuations. The gas disk with a FWHM size of 530AU $\times$ 330AU
(P.A.~$\sim$ 66.7$^{\circ}$), derived from Gaussian fitting of the data of the lowest resolution data set,
appears to be larger than the dust disk. By contrast with the dust disk, the gas disk exhibits a peak at the
stellar position (right panel of Figure \ref{Fig 3}). Since $^{12}$CO(3-2) is optically thick and the dust
continuum at 850 $\micron$ is optically thin, this suggests that the central peak in the gas disk is a
temperature effect. Indeed, the gas disk also shows a central depression in optically thinner lines, such as
$^{13}$CO and C$^{18}$O (P05), consistent with this interpretation.

There is a secondary peak in the $^{12}$CO(3-2) intensity map coincident with the NE peak in the dust disk,
which possibly traces the inner spiral arm (see the right panel of Figure \ref{Fig 3}). The $^{12}$CO(2-1)
emission observed by P05 does not peak at this position. This is most probably due to insufficient angular
resolution of their $^{12}$CO(2-1) map. When we convolve our $^{12}$CO(3-2) map with the same beam size of the
$^{12}$CO(2-1) map ($2''\!\!.0\times~1''\!\!.6$), the secondary peak disappears and the resultant map looks
similar to the PdBI map.

As shown in the left panel of Figure \ref{Fig 3}, the largest
velocity gradient is along the major axis, consistent with the
pattern of a rotating disk. Therefore the molecular gas emission is
interpreted as a large rotating disk as previous millimeter
observations have shown \citep{Mannings:1997, Corder:2005,
Pietu:2005}. Although the bulk motion of the gas disk looks
agreeable with rotation, there are also velocity gradients along the
minor axis. In particular, at the south end of the minor axis, there
is an abrupt change in velocity, suggestive of non-circular motion.

\begin{figure}
\begin{center}
\includegraphics[width=\textwidth]{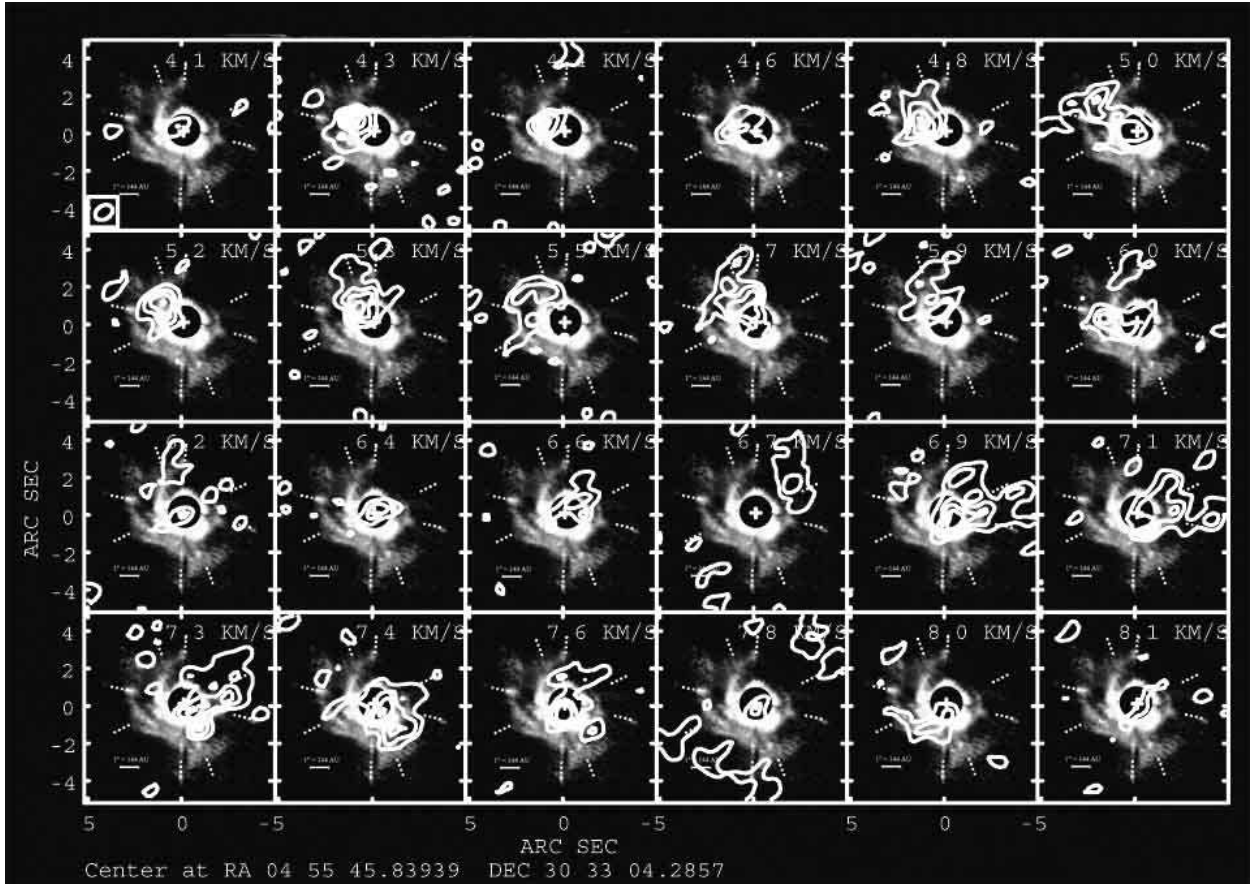}
\caption[$^{12}$CO (3-2) channel maps]{CO channel maps (in contours)
with natural weighting. Contour spacing is 2$\sigma$ (or 0.5Jy
Beam$^{-1}$). The background in every velocity channel is the
near-IR image from Subaru (F04) Note that there is significant CO
emission beyond the near-IR image. Note also that some velocity
channels trace the spiral-like structures. \label{Fig 2}}
\end{center}
\end{figure}

Figure \ref{Fig 2} shows channel maps of the $^{12}$CO(3-2) emission superposed on the Subaru image. We found a
general velocity gradient along the major axis as seen in the MOM 1 map. However, the maps do not show a simple
``butterfly'' diagram as seen in the $^{13}$CO maps of OVRO \citep{Corder:2005} and PdBI (P05) at comparable
angular resolution, or the $^{12}$CO maps at lower transitions at lower angular resolution. We found that in
several velocity channels - mainly from 4.8 km s$^{-1}$ to 6.2 km s$^{-1}$ - the CO emission traces the inner
spiral arm inferred by F04 from scattered light. The CO emission in the 8.0 km s$^{-1}$ channel coincides with
the outer arm, but it is only of 4 sigma significance and is the only channel tracing the outer arm. There are
also many channels, most notably at 5.0 km s$^{-1}$, 6.7 km s$^{-1}$ to 7.3 km s$^{-1}$, and 7.8 km s$^{-1}$,
where the emission extends well outside of the near-IR images. These features suggest that the CO emission
traces extended structure, presumably the envelope or the outer region of a much larger CO disk, as well as a
more compact disk. A simple model to better delineate which parts of the emission in the channel maps show
non-circular or non-Keplerian rotation will be discussed in the next section.

\section{Discussion}
\subsection{Disk Mass and SED}
Based on single dish observations, \cite{Acke:2004} deduced the power-law index $n$ in $\lambda F_{\lambda}
\propto \lambda ^{-n}$ for the spectral energy distribution (SED) of AB Aur  to be as high as $4.32$, suggesting
a particle size distribution dominated by relatively small grains and therefore little grain growth. However,
the flux of dust continuum emission we observed is only 65\% of the single dish measurement. This implies that
the single dish data are also sensitive to the large scale envelope, where there may be indeed relatively little
if any grain growth.

We derived a spectral index $n = 3.70 \pm 0.21$ by fitting the total flux density from interferometers with a
typical angular resolution of 1$''$ at 850 $\micron$ (SMA), 2 mm (NMA; Ohashi et al.~in preparation), 1.4 mm and
2.8 mm (PdBI, P05). The obtained value of $n$ is only 3 $\sigma$ different from the single dish result.
Therefore there may be more grain growth than \cite{Acke:2004} thought, even though this number still suggests a
relatively primitive disk \citep[e.g.~see][]{Pietu:2005}. Note that the power-law index is averaged over the
entire disk, and given its observed structure at millimeter and submillimeter wavelengths is weighted toward the
outer region. There may still be significant grain growth further in. We infer a disk mass of $0.0075 \pm 0.0030
\Msun$, assuming the emissivity $K(\nu) = 0.1(\nu / 10^{12} Hz)^{\beta}$ cm$^{2}$ g$^{-1}$
\citep{Beckwith:1990}, a dust temperature of 26K derived from the average intensity of the CO (3-2) emission,
and a gas-to-dust mass ratio of 100. Assuming the dust continuum emission to be completely optically thin,
i.e.~$ n = \beta + 3$ \citep{Beckwith:1990}, the dust emissivity index $\beta$ was estimated to be 0.70 from the
spectral index \textit{n}. The disk mass is consistent with the OVRO observations
\citep[0.009$\Msun$,][]{Corder:2005}, but is smaller than the mass P05 obtained by a factor of 2. Note that the
$\beta$ Pi\'etu et al.~derived ($\beta \sim 1.4$) from their 1.4mm and 2.8mm dust maps is without the optically
thin assumption and hence is larger. If we use the same $\beta$ as Pi\'etu et al.~used, the derived disk mass is
consistent (0.016$\Msun$ $\pm$ 0.006$\Msun$).

\subsection{Kinematics of the Spiral Arms}

\begin{figure}
\begin{center}
\includegraphics[width=\textwidth]{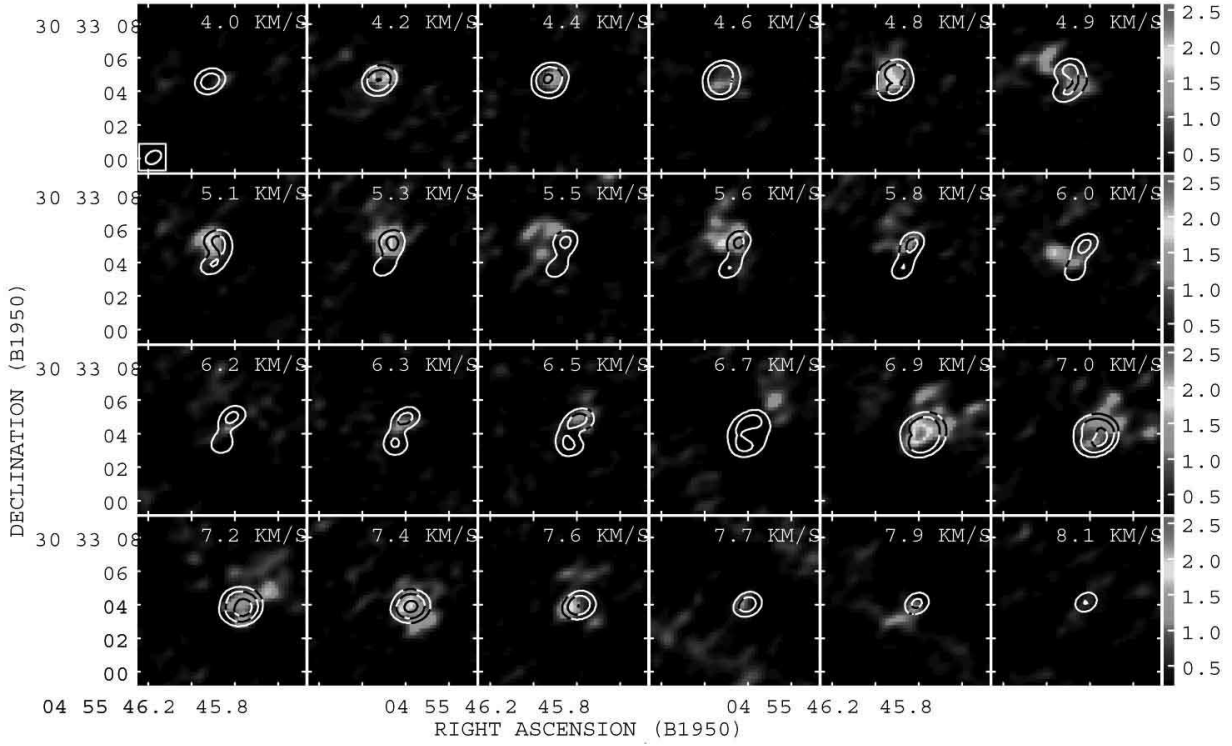}
\caption[Observed and modeled CO(3-2) channel maps]{Observed and modeled CO(3-2) channel maps. The contours are
the 10\%, 30\%, and 70\% intensity levels of the modeled CO emission from a power-law distributed ($exponent =
-0.47$) intensity Keplerian disk. The model contours appear to be not exactly symmetric because it has been
convolved with the same beam as the observation, with a PA of $\sim$ -58$^{\circ}$. Color maps are the observed
CO emission, starting from 2$\sigma$ ($\sim$0.5Jy Beam$^{-1}$).\label{Fig 4}}
\end{center}
\end{figure}

P05 has made a $\chi^2$ model fitting for the disk from their $^{12}$CO/$^{13}$CO observations at an angular
resolution of around 1$''$. The deduced velocity laws all deviate from Keplerian motions. The radial velocity
profile is shallower than Keplerian from $^{13}$CO, but steeper from $^{12}$CO. Their $^{13}$CO(2-1) results are
probably the best data on which to base such a model, due to good signal-to-noise ratio, and the highest angular
resolution currently available.  In addition, being optically thin allows this line to sample the entire depth
of the disk. Given the limitation of our data quality, our goal here is not to derive a dynamical model of the
disk. Instead we intend to illustrate that the motion of the disk - in particular the part that traces the main
spiral arm - clearly deviates from Keplerian. By comparing our data with a simple disk model with a power-law
distributed intensity and Keplerian rotation, we attempt to identify which parts of the emission in the channel
maps exhibit non-Keplerian motion. We use the best fit parameters derived from the $^{13}$CO(2-1) observations
done by P05 to define our model: a central mass of 2.2 $\Msun$ and an inclination angle (\textit{i}) of
42$^{\circ}$. The outer radius for the disk is assumed to be $530$ AU, as inferred by our $^{12}$CO data.  We
also fit a power-law distribution of $^{12}$CO(3-2) brightness temperature of $T = T_{100} (r/100AU)^{-0.47\pm
0.02}$, $T_{100} \sim 23.4 K$. Figure \ref{Fig 4} shows the observed and the modeled channel maps. The
observational data, plotting from 2$\sigma$ ($\sim$0.5Jy), are shown in color. The white contours indicate the
10\%, 30\%, and 70\% intensity levels of emission of the model. Since the dynamical range of every velocity
channel is never over 10, emission that falls within the outermost contours is consistent with the
interpretation that the gas motion follows Keplerian rotation law. On the other hand, emission that falls
outside this boundary must deviate significantly from the gas motion defined by our model disk.

As shown in Figure \ref{Fig 4}, most of the observed emission falls inside the boundary. This suggests that the
bulk motion of the disk is consistent with Keplerian, and thus is consistent with being gravitationally bound by
the central star. However, there are excess emission components which can not be explained by the Keplerian
kinematics, especially in the velocity channels from 5.5~km~s$^{-1}$ to 6.0~km~s$^{-1}$. Moreover, the excess
emission components are spatially consistent with the inner spiral arm. This is where the emission coincides
with the inner spiral arm. The dynamical ranges of these four channels are no more than 3, suggesting that the
deviation from Keplerian is probably even more substantial than this simple model comparison suggests. If we use
a power law radial dependence for the velocity with an index of $-0.42$ which P05 derived from the
$^{13}$CO(2-1) observation, the inner arm still falls outside the model boundary. The inner spiral arm seen in
the scattered light image of F04 is therefore a physical arm which has different kinematics as compared with the
bulk of the disk. There is also deviation from Keplerian motion in the 8.0 km s$^{-1}$ channel which might trace
the outer arm. Note that even though the emission falls inside the outermost contours, the emission peaks often
do not coincide with the predicted positions of emission maxima (Figure \ref{Fig 4}). We made a similar
comparison with the $^{13}$CO data of P05. The bulk motion of the emission falls inside the contours, and the
emission maxima are consistent with the positions predicted by the model. This implies that what we are really
seeing in $^{12}$CO is largely a temperature effect. The observed differences strongly suggest that our
$^{12}$CO(3-2) map preferentially traces elevated gas temperatures at or close to the surface of the disk.

\begin{figure}
\begin{center}
\includegraphics[width=0.8\textwidth]{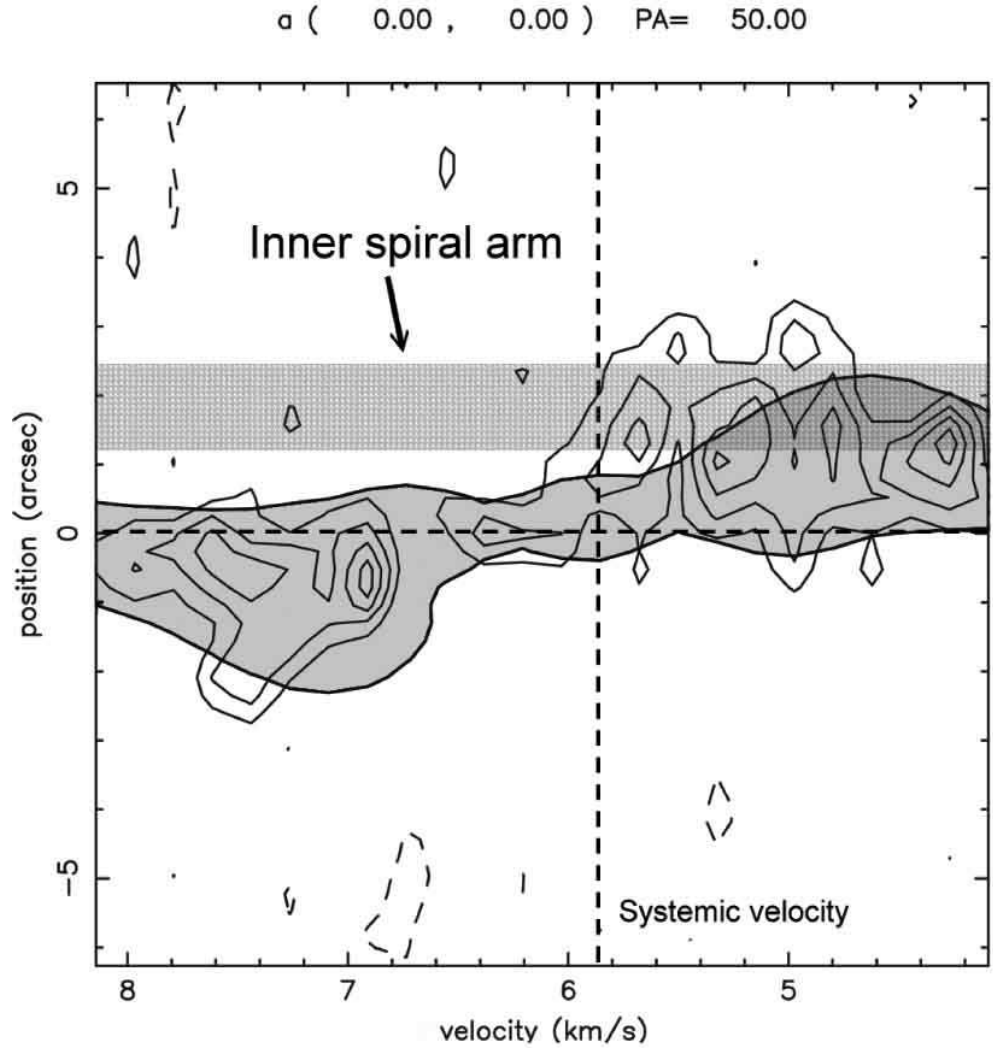}
\caption[Position-Velocity diagram along
P.A.~$\sim$50$^{\circ}$]{Position-velocity diagram along
P.A.$\sim$50$^{\circ}$. The emission (in contours with -2$\sigma$,
2$\sigma$, 4$\sigma$, 6$\sigma$, and 8$\sigma$ intensity) lying
outside the gray-shaded area (10\% peak intensity) cannot be
explained by the Keplerian motion. The hatched area indicates the
position of the inner spiral arm. The systemic velocity is
$\sim5.84~km~s^{-1}$. The position-velocity diagram has not been
smoothed across the cut.\label{Fig 19}}
\end{center}
\end{figure}

\begin{figure}
\begin{center}
\includegraphics[width=0.5\textwidth]{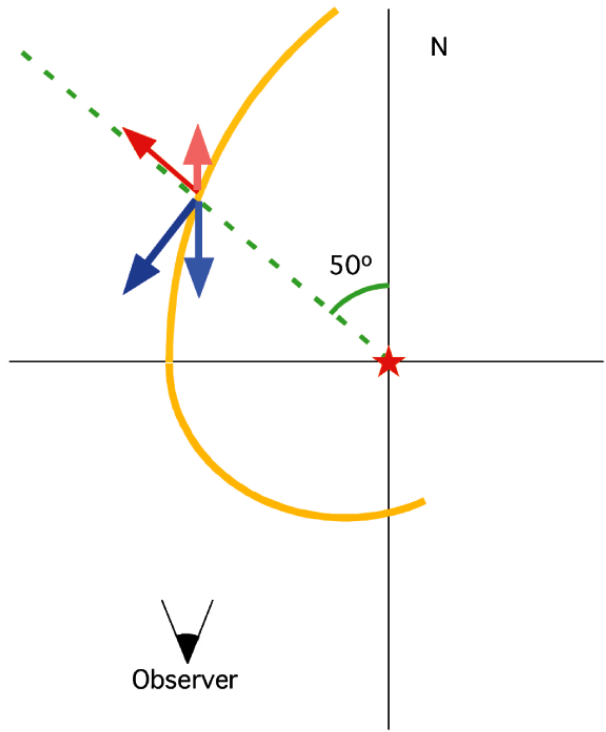}
\caption[Streaming motion of the spirals]{Diagram illustrating the motion on the spirals. N denotes the north.
The yellow curve indicates the inner spiral arm. The deep blue arrow represents the direction of the Keplerian
motion of the material going through the spiral. The red arrow indicates the outward radial motion. The light
blue and pink arrows are the components along the line-of-sight of the Keplerian motion and the radial outward
motion, respectively. \label{Fig 18}}
\end{center}
\end{figure}

We show a position-velocity (P-V) diagram in Figure \ref{Fig 19} to illustrate the position of the excess
emission more clearly. Note that the cut along the major axis (P.A.$\sim67^{\circ}$) does not pass through the
bulk of the emission tracing the main arm, and the non-Keplerian features are not prominent along this
particular cut. Therefore, we have selected another cut with P.A.~$\sim50^{\circ}$, which passes through the
bulk of the emission in the arm, and compared the P-V diagram of the Keplerian model along the same cut. The P-V
diagram clearly exhibits anomalous emission which can not be explained by the Keplerian motion at the position
of the inner arm. This is consistent with what we found in the channel maps. Most importantly, the emission of
the deviation is red-shifted with respect to the expected Keplerian motion. This red-shifted motion can be
explained by possible streaming motion along or across the spiral arms. Spiral arms are the product of
nonaxisymmetric gravitational potential (either due to the presence of a planet or due to self-excited
gravitational instability), and hence would be expected to show radial motion in addition to the circular
Keplerian motion. In the case of AB Aur, based on the Subaru image, the bright southeast side is the near side.
Together with the information from our MOM1 map, the spiral arms are trailing, and the motion across the arms is
radially outward. At P.A$\sim50^{\circ}$, as shown in Figure \ref{Fig 18}, the outward streaming motion on the
spiral arm contributes an additional red-shifted component in the line-of-sight velocity, which accounts for the
anomalous emission. However, on the premise that our model is correct, if the excess emission is indeed the
result of outward streaming motion, the cause of the large magnitude outward motion ($\sim$ 1 km s$^{-1}$, from
comparing the peak position of a Keplerian disk) may be enigmatic. We should be careful that the above statement
provides only a qualitative argument about how the materials around the spiral arms deviate from Keplerian
motion, and needs to be verified by a more detailed model.

We also take notice of other possibilities to interpret the
non-Keplerian kinematics. For example, in the case of HL Tau,
various motions of the circumstellar gas that may happen around an
embedded young system, such as outflows \citep{Cabrit:1996} and
infalls \citep{Hayashi:1993} in addition to the large-scale
Keplerian rotation \citep{Sargent:1987}, have been proposed to
explain the complex kinematics. In AB Aur case, there is no evidence
for any ongoing or recent outflows, or any ionized jet seen at
centimeter wavelengths, or any evidence for molecular outflows.
Although AB Aur is still embedded in a large scale envelope, we
expect that the interferometer will filter out the extended
structure larger than 13'' (radius $>$900~AU). In addition, the fact
that the CO (3-2) line is a better tracer of the warmer and denser
region of the molecular gas, suggests that the non-Keplerian motions
may be mainly from the disk itself. The anomalous emission coincides
so well with the inner spiral arm, that the emission could come from
the spiral arm.

\subsection{Physical Nature of the Spiral Arms}
The disk of AB Aur exhibits great complexity in multi-wavelength studies by various authors. The SMA observation
show clearly asymmetric structures and disturbed kinematics, which may interpret as a spiral arm in the high
resolution $^{12}$CO(3-2) maps.  In $^{13}$CO observations with better resolution and higher dynamical range
done by PdBI (P05), the channel maps resemble a butterfly diagram, and the signature of a spiral arm is not as
clearly seen as in $^{12}$CO(3-2). Although in some velocity channels there are extended structures which may be
contributed by the inner spiral arm. On the other hand, in the $^{12}$CO(3-2) channel maps, the emission
patterns are nothing like the standard butterfly diagram. Note that $^{12}$CO(3-2) is the higher transition of
the main CO isotope, and it traces the emission closer to the warmer disk surface heated by the central star,
while the optically thin $^{13}$CO(2-1) and $^{13}$CO(1-0) lines trace emission all the way to the cold midplane
of the disk. Hence, the overall discrepancy of the channel map patterns, as well as the spiral arms in
$^{12}$CO(3-2) may be due to temperature enhancement rather than column density effect. Moreover, this could be
a ``surface'' effect. The $^{12}$CO emission is more optically thick and weighted towards the surface layers,
whereas the $^{13}$CO emission is more weighted towards deeper cooler layers. The lack of dust coinciding with
the spiral arm also supports this argument (right panel in Figure \ref{Fig 3}).

From the optically thin dust continuum observation at 850 $\micron$,
we also found a central depression and asymmetric local enhancement
in density that forms a ring-like structure which may be the inward
extension of the inner spiral, as P05 also suggest in their 1.4 mm
map. This ring-like structure, however, should be due to the column
density effect. The NE peak in the 850 $\micron$ dust continuum map
is absent in the 1.4 mm dust map. This suggests that the SED at this
position may be quite different from other parts of the disk. Grain
growth or complexity in chemical composition resulting in different
dust emissivity may be the cause.

In summary, the observational results may be interpreted as a
slightly higher density distribution at the spiral position, where
the density contrast against the rest of the disk is not high enough
to be picked out in optically thin lines. In addition, the
temperature at the position of the spiral arm may be higher than the
rest of the disk.  Gas in the arms may be compressed and heated.
There may also be a larger scale height at the spiral position so
that material therein near the disk surface will receive more direct
radiation from the central star, and become warmer. Hence, optically
thick lines such as $^{12}$CO(3-2), can be more sensitive to the
spiral structure.

\subsection{Excitation Mechanisms}
Both F04 and P05 discussed the possibility of exciting the
spiral-like structure by the self-gravity of the disk. With the
surface density derived from our dust continuum observation, we can
estimate the Toomre's Q parameter:
\[
Q \sim \frac{c_s \kappa}{\pi G \Sigma}~,
\]
where c$_s$ is the sound speed, $\kappa$ is the effective angular
velocity which is $\sim$ $\Omega$ (Keplerian angular veolcity), and
$\Sigma$ is the surface density. We can thus evaluate the disk
stability at least in the inner 200AU region of the dust disk where
the emission extends out to the inner spiral arm and further. We fit
a power-law distribution of CO (3-2) brightness temperature of T =
T$_{100}$ (r/100AU)$^{-0.47\pm0.02}$. As a zeroth order estimation,
we assume that the dust temperature is coupled to that of the gas.
At the radius of 150AU, with a surface density of 2.74 $\pm$ 1.04 g
cm$^{-2}$ (K $\sim$ 0.05 cm$^2$ s$^{-1}$; gas-to-dust ratio of 100)
and $T_{gas}\sim T_{dust}\sim 20.3\pm6.6 K$, we derived a Q of 9.5
$\pm$ 4.0. The Q value we derive is about 2$\sigma$ larger than
unity, the value at or below which the disk is gravitationally
unstable. Note that P05 also found a Q value $\sim$ 11 at 100AU with
a much smaller uncertainty, certainly higher than unity. Therefore
the gravitational instability mechanism is not favored by current
observations, even though we also note that the disk mass estimation
we performed still has large ambiguity because of the missing flux
and uncertainty of the dust emissivity.

\cite{Pietu:2005} discussed the possibility of excitation by a
coeval stellar companion. A number of optical/infrared studies place
stringent mass limits on any such stellar companion (see P05).
Instead, the spiral arms may be excited by a giant planet or
planets. A giant planet inside a circumstellar disk can excite
density waves outward, and hence produce a prominent one-arm spiral
from its position. Consequently, if there is a (forming) giant
planet at about several tens AU to a hundred AU away from AB Aur,
the inner arm can be formed and maintained by the density waves it
excites. A giant planet with an orbit on a different plane from the
circumstellar disk may also excite bending waves that puff up the
spiral so that at the spiral position the tilted surface can receive
more stellar radiation and become warmer, and hence can be detected
by the higher transitions of CO. The relatively large inner (column
density) depression in the dust and gas disk provides additional
evidence for a giant planet. For example, the depression of emission
in the central dust disk resembles the central cavity of Vega-type
stars: the inner radii of the ring-like structure around AB Aur and
the asymmetric structure around Vega-type stars are of the same
order ($\sim$70 - 80AU). This phenomenon can also be associated with
the existence of a giant planet at about several tens of AU to
$\sim$ 100 AU to clear out a cavity. Note that the central
depression of dust continuum does not have to imply a cavity (hole).
For example, a similar structure can be created with the presence of
a spiral arm or a ring which enhances local column density further
out in the disk.

\section{Summary}
Our SMA observations of the Herbig Ae star AB Aur, on the dust
continuum emission at 850$\micron$ and the $^{12}$CO(3-2) line
emission, resolve its circumstellar disk at an angular resolution of
$1''\!\!.0 \times 0''\!\!.7$. The disk is much more complicated than
ordinary Keplerian disks observed around other Herbig Ae stars. The
observations reveal that:
\begin{enumerate}
\item The dust emission exhibits an asymmetric local density enhancement that
forms a ring-like structure with a central depression. The derived disk mass is 0.0075 - 0.016 $\Msun$, for a
dust emissivity index of 0.7 - 1.4. This is consistent with the results of \cite{Corder:2005} and
\cite{Pietu:2005}.
\item In addition to the gas motion delineated by the best fit disk model of P05, we also
found gas components associated with the spiral arms identified in
the near-IR image. Their motion deviates from the Keplerian
kinematics expected from the stellar mass of AB Aur. Streaming
motion along and across the spiral arm may account for the
deviation.
\item The radial outward motions of the spiral arms as well as a central cavity within the
disk, suggest the possible existence of a giant planet forming in the disk. This is consistent with a relatively
small mass for the disk, which is probably not large enough to excite spiral arms due to gravitational
instability.

\end{enumerate}

Acknowledgement: We thank M. Momose, F. Shu, and C. Yuan for fruitful discussion. N.O.~was supported in part by
NSC grant NSC93-2112-M-001-042.
\pagebreak


\bibliographystyle{astron}
\bibliography{mnemonic,abaur}
\end{document}